\documentclass[12pt,a4paper,oneside,onecolumn]{article}
\usepackage[cp1252]{inputenc}
\usepackage[english]{babel}
\usepackage{amsmath}
\usepackage{amssymb}
\usepackage{graphicx}
\usepackage{setspace}
\usepackage[a4paper,left=3cm,right=2cm,top=2cm,bottom=2cm]{geometry}
\usepackage[square,comma,numbers,sort&compress]{natbib}
\usepackage{sectsty}
\usepackage[symbol]{footmisc}
\usepackage[labelsep=period]{caption}
\allsectionsfont{\large\centering}
\makeatletter
\def\@seccntformat#1{\csname the#1\endcsname.\quad}
\makeatother
\begin{document}
\begin{flushleft}
\textit{\large Astronomy Reports, 2010, Vol. 54, No. 4, pp. 355---366.}
\footnote{Original Russian Text published in Astronomicheskii Zhurnal, 2010, Vol. 87, No. 4, pp. 397---409.}
\end{flushleft}
\vspace{\baselineskip}
\begin{center}
\begin{Large}
\textbf{The Appearance of a Radio-Pulsar Magnetosphere from a Vacuum with a Strong Magnetic Field.\linebreak Accumulation of Particles}
\end{Large}
\\ \bigskip
\textrm{\large Ya. N. Istomin$^1$ and D. N. Sob'yanin$^2$}
\\ \bigskip
\textit{$^1$Lebedev Physical Institute, Russian Academy of Sciences,\\
53 Leninskii pr., Moscow, Russia}
\\ \medskip
\textit{$^2$Moscow Institute of Physics and Technology (State University),\\
Dolgoprudnyi, Moscow oblast, Russia}
\\ \bigskip
Received June 30, 2009; in final form, October 19, 2009
\\ \bigskip
\begin{large}
\textbf{Abstract}
\end{large}
\end{center}
\vspace{-0.3cm}
\par\textrm{The accumulation of electrons and positrons in the vacuum magnetosphere of a neutron star with a surface magnetic field of $B\sim10^{12}$~G is considered. It is shown that particles created in the magnetosphere or falling into the magnetosphere from outside undergo ultra-relativistic oscillations with a frequency of $10-100$~MHz. These oscillations decay due to energy losses to curvature radiation and bremsstrahlung, with their frequencies reaching $1-10$~GHz. Simultaneously, the particles undergo regular motion along the force-free surface along closed trajectories. This leads to the gradual accumulation of particles at the force-free surface and the formation of a fully charge-separated plasma layer with a density of the order of the Goldreich–Julian density. The presence of a constant source of electron–positron pairs in the magnetosphere due to the absorption of energetic cosmic gamma-rays leads to the growth of this layer, bringing about a rapid filling of the pulsar magnetosphere with electron–positron plasma if the pair-creation multiplication coefficient is sufficiently high.}
\newpage
\section{INTRODUCTION}

The generation of electron–positron plasma in the magnetosphere of a rotating,magnetized neutron star has been well studied \citep{Sturrock1971,RudermanSutherland1975,GoldreichJulian1969}. This has been used as a basis for developing the physics of a stationary pulsar magnetosphere filled with plasma whose density appreciably exceeds the so-called Goldreich–Julian density \citep{GoldreichJulian1969}, $n_{GJ}=|\mathbf{\Omega}\cdot\mathbf{B}|/2\pi ce$, corresponding to stationary rotation of the magnetosphere right out to the light cylinder, $R_L=c/\Omega$. Here, $\Omega$ is the angular velocity of the star’s rotation, $c$ the speed of light, and $e$ the positron charge. However, it is difficult to understand what mechanisms are generating the radio emission of pulsar magnetospheres, where the plasma is created, based on observations of radio pulsars operating in a steady-state regime. Tracing the dynamics of the development of the emission at different frequencies could prove important for our understanding of physical processes occurring in radio pulsar magnetospheres \citep{GurevichIstomin2007}.

Moreover, many observations of non-stationary radio pulsars have recently appeared. These concern first and foremost so-called switching radio pulsars, whose radio emission is observed only during certain time intervals appreciably exceeding the rotational period of the star, such as the pulsars PSR B1931+24 \citep{KramerEtal2006} and PSR J1832+0029 \citep{Kramer2008}. In addition to switching pulsars, a group of so-called nulling pulsars has long been known, which likewise display no radio emission during certain intervals, but with these being not as regular as those for switching pulsars \citep{WangEtal2007}. Another group of non-stationary radio sources has recently been observed: rotating radio transients (RRATs), which are sporadically flaring radio sources. The phases are preserved during these flares, and the corresponding measured periods are characteristic of ordinary radio pulsars \citep{McLaughlinEtal2006}. There is no doubt that these are also rotating neutron stars. In our view, all these sources exhibit a non-stationary generation of plasma in neutron-star magnetospheres. It is therefore important to understand how the magnetosphere of a rotating, magnetized neutron star is filled with plasma, which is the reason for the operation of radio pulsars.

Together with \citep{Istomin1Sob'yanin2008}, the current study is concerned with an initial investigation into the ``ignition'' of the magnetospheres of neutron stars---the dynamics of the filling of the vacuum magnetosphere with electrons and positrons created in the magnetosphere. In \citep{Istomin1Sob'yanin2008}, we analyzed the dynamics of the motion of particles in the vacuum magnetosphere of a rotating neutron star possessing a strong magnetic field ($\geqslant10^{12}$~G). Here, we consider the deceleration, capture, regular motion, and accumulation of particles near the force-free surface in the magnetosphere---the surface where the electric field along the magnetic-field lines vanishes.

\section{DECELERATION OF THE PARTICLE MOTION}

In our study of the motion of charged particles \citep{Istomin1Sob'yanin2008}, we showed that particles approaching the force-free surface begin to undergo oscillations. During these oscillations near the force-free surface, the charged particles lose energy to radiation. Let us consider the time dependence of the rate of decrease of a particle’s energy. The particle’s energy is characterized by the Lorentz factor possessed when the particle crosses the force-free surface, which we denote $C$. Our main interest is the variation of the energy in the ultra-relativistic case, when $C\gg1$; in the non-relativistic case, $C\simeq1$, and this quantity remains essentially constant.

We obtain from the Dirac–Lorentz equation
\begin{equation}
\label{DLEforUltrarelativisticOscillations}
\frac{dC}{dt}=\frac{2}{3}\alpha\gamma\!\left[\frac{d^2\gamma}{dt^2}
-\frac{\gamma^3}{\rho^2}\right],
\end{equation}
where $\gamma$ is the particle’s Lorentz factor, $\rho$ the radius
of curvature of the magnetic-field line along which the particle moves, and $\alpha=e^2/\hbar c\approx1/137$ the fine-structure constant. Here and below, we use dimensionless units, measuring the strengths of the electric and magnetic fields in units of the so-called critical field $B_{cr}=m_e^2c^3/e\hbar\approx4.4\times10^{13}$~G, the particle velocity in units of the speed of light $c$, the particle charge in units of the positron charge $e$, the particle mass in units of the electron mass $m_e$, the particle
energy in units of the electron rest-mass energy $m_e c^2$, all distances in units of the Compton wavelength of the electron $^-\!\!\!\!\lambda=\hbar/m_e c\approx3.86\times10^{-11}$~cm, and all times in units of $^-\!\!\!\!\lambda/c$. Note that, in these units, $1000\text{ km}\approx2.6\times10^{18}$ and $1\text{ s}\approx7.8\times10^{20}$. Then, $C=\gamma+\omega^2(x'_1)^2/2=1+\omega^2A^2/2$, where $\omega$ is the frequency of the non-relativistic oscillations, $x'_1$ the current longitudinal coordinate of the particle along the magnetic-field line, and $A$ the amplitude of the oscillations. The frequency of the non-relativistic oscillations is given by
\begin{equation}
\label{nonrelativisticOscillationsFrequency}
\omega^2=-\frac{\mathbf{b}_0\cdot\nabla(\mathbf{E}_0\cdot\mathbf{B}_0)}{\mathbf{B}_0},
\end{equation}
where $\mathbf{E}_0$, $\mathbf{B}_0$, and $\mathbf{b}_0=\mathbf{B}_0/B_0$ are the electric and magnetic vectors and a unit vector along the magnetic-field line at the point where it intersects the force-free surface. To order of magnitude, the frequency of the non-relativistic oscillations is $\omega\sim\sqrt{B/R_L}$, which corresponds to the frequency $\nu=\omega/2\pi\sim1{-}10$~GHz.

If we neglect the radiation-reaction force, which formally corresponds to $\alpha=0$, then $C=\mathrm{const}$. In this case, $C$ will be an exact first integral. However, generally speaking, there is a radiative reaction, and the particle will lose energy. This leads to a decrease in $C$. The rate of decrease of $C$ will not be constant over times of the order of the period of the particle motion, because it depends on the specific Lorentz factor of the particle $\gamma$, whose value must obviously change over one-quarter of the oscillation period from $1$ at the turning point where $x'_1=A$ to $C\gg1$ at the point $x'_1=0$. However, this will not hinder our investigation of the variation of the particle energy, as long as $C$ itself varies little over the oscillation period:
\begin{equation}
\label{adiabaticCondition}
\frac{dC}{dt}\ll\frac{C}{T}.
\end{equation}
Satisfaction of this condition of applicability of the adiabatic approximation means that we can accurately average \eqref{DLEforUltrarelativisticOscillations} over the oscillation period. We must average expressions that depend on the particle coordinate $x'_1$ not formally, but instead using the explicit relation $x'_1(t)=A(-1)^{h(z)}\left(z-2h(z)\right)$ for the time dependence of the coordinate of the oscillating particle, which is exact when $C\gg1$. The function $h(z)$ is given by $h(z)=\lfloor(z+1)/2\rfloor$, where $z=t/A$ and $\lfloor y\rfloor$ denotes the integer part of $y$.

Let us suppose we have a function $f$, which depends on the distance from the equilibrium point, with the remaining parameters being arbitrary, with the only specification being $f=f(|x'_1(t)|)$. It can then easily be shown that an average over the period $T$ can be replaced by an average over one-quarter of the period $A$. If we express $f$ as a function of $z=t/A$ using the transform $f=\tilde{f}(z)=f(|x'_1(Az)|)$, the mean of $f$ over the oscillation period will be given by
\begin{equation}
\label{averagingFormula}
\langle f\rangle=\int\limits_0^1\tilde{f}(z)\,dz.
\end{equation}
The angular brackets denote the mean of a function over the oscillation period. To average the right-hand part of \eqref{DLEforUltrarelativisticOscillations}, we must know the Lorentz factor $\gamma$:
\begin{equation}
\label{gammaFactor}
\gamma\simeq C\left[1-(z-2h(z))^2\right]\!.
\end{equation}
Formula \eqref{gammaFactor} was obtained using the relation $C\simeq\omega^2A^2/2$. Further, we neglected unity compared to $C$ in the expression for $\gamma$, which does not affect the averaging results, since $C\gg1$. After averaging \eqref{DLEforUltrarelativisticOscillations} using \eqref{averagingFormula}, we obtain the differential equation for $C$
\begin{equation}
\label{averagedDLEforUltrarelativisticOscillations}
\frac{dC}{dt}=-\frac{4}{9}\alpha\omega^2C
\left[1+\frac{64}{105}\,\frac{C^3}{\omega^2\rho^2}\right].
\end{equation}
The exact solution of \eqref{averagedDLEforUltrarelativisticOscillations} has the form
\begin{equation}
\label{cSolution}
\varepsilon(t)=\left[\left(1+\varepsilon_0^{-3}\right)e^{3t/\tau_{d}}-1\right]^{-1/3}.
\end{equation}
where $\varepsilon(t)=C(t)/C_{curv}$ is the normalized first integral and
$\varepsilon_0=\varepsilon(0)$ its value at the initial time $t=0$. Here, we have also introduced
\begin{equation}
\label{Ccurv}
C_{curv}=\frac{105^{\,1/3}}{4}(\omega\rho)^{2/3}\approx1.2\:(\omega\rho)^{2/3}
\end{equation}
and the decay time constant
\begin{equation}
\label{tauD}
\tau_{d}=\frac{9}{4}\,\frac{1}{\alpha\omega^2}.
\end{equation}

The physical meanings of $C_{curv}$ and $\tau_d$ are not difficult to understand based on the behavior of the solution \eqref{cSolution} for various initial values $\varepsilon_0$. Let first $\varepsilon_0\ll1$. The solution of \eqref{averagedDLEforUltrarelativisticOscillations} is then given by
\begin{equation}
\label{undercurvSolution}
\varepsilon(t)=\varepsilon_0e^{-t/\tau_d}\qquad(C_0\ll C_{curv}).
\end{equation}
We can see that the case $C_0\ll C_{curv}$ corresponds to leaving out the term in parentheses, proportional to $(C/C_{curv})^3$ in \eqref{averagedDLEforUltrarelativisticOscillations}. However, this means that the energy losses of the charged particle to curvature radiation are small compared to losses due to the term $\propto\gamma\,d^2\gamma/dt^2$ in \eqref{DLEforUltrarelativisticOscillations}. These energy losses can also be associated with bremsstrahlung losses, due to the oscillatory motion of the particle. These will differ from zero only if the particle energy varies with time. In contrast, the curvature losses depend only on the radius of curvature of the particle trajectory and the particle energy. There is clearly no dependence of the intensity of curvature radiation on time derivatives of the energy of the charged particle. Thus, when $C_0\ll C_{curv}$, the particle energy decreases exponentially with a decay time constant $\tau_d$ \eqref{tauD}, and the losses themselves occur mainly due to bremsstrahlung.

Now let $\varepsilon_0\gg1$. Obviously, in this case, the most interest is presented by the behavior of $C$ on relatively short times $t$, since $C$ decreases with time and will become smaller than $C_{curv}$ at sufficiently long times, so that the decay takes on an exponential character. At small times $t$, we have
\begin{equation*}
\label{curvSmallTimeSolution}
\varepsilon(t)=\varepsilon_0\left(1-\frac{t}{\tau_{p}}\right)
\qquad(C_0\gg C_{curv},\text{ }t\ll\tau_{p}),
\end{equation*}
where
\begin{equation}
\label{tauP}
\tau_{p}=\frac{\tau_d}{\varepsilon_0^3}.
\end{equation}
We can see that, if $C_0\gg C_{curv}$, there will be a linear decrease in the particle energy on times appreciably less than $\tau_p$. When $t$ is comparable to $\tau_p$, the dependence acquires the form
\begin{equation*}
\label{curvSmallIntermediateTimeSolution}
\varepsilon(t)=\varepsilon_0\left(1+3\frac{t}{\tau_{p}}\right)^{-1/3}
\qquad(C_0\gg C_{curv},\text{ }t\sim\tau_{p}).
\end{equation*}
With further growth in $t$, the decay becomes a power law:
\begin{equation*}
\label{curvIntermediateTimeSolution}
\varepsilon(t)=\left(3\frac{t}{\tau_d}\right)^{-1/3}
\qquad(C_0\gg C_{curv},\text{ }\tau_{p}\ll t\ll\tau_d).
\end{equation*}
Here, the meaning of the time $\tau_p$ becomes clear, as the time during which the decrease of the energy of the charged particle, which is initially linear, acquires a power-law character. On times of the order of $\tau_d$, the dependence again changes:
\begin{equation*}
\label{curvIntermediateLargeTimeSolution}
\varepsilon(t)=\left(e^{3t/\tau_d}-1\right)^{-1/3}
\qquad(C_0\gg C_{curv},\text{ }t\sim\tau_d).
\end{equation*}
On still longer times, the dependence becomes exponential:
\begin{equation}
\label{curvLargeTimeSolution}
\varepsilon(t)=e^{-t/\tau_d}
\qquad(C_0\gg C_{curv},\text{ }t\gtrsim\tau_d).
\end{equation}

In contrast to the case \eqref{undercurvSolution} considered earlier, when $C_0\gg C_{curv}$, the particle’s energy losses are due primarily to curvature radiation. This process dominates on times $t\lesssim\tau_d$ and has a power-law character. When $t\gtrsim\tau_d$, the decrease in the power energy becomes exponential, indicating a transition to a regime in which the main contribution to the energy losses is made by bremsstrahlung associated with the acceleration experienced by the particle during its oscillatory motion. It is clear that this transition itself occurs on time scales of the order of $\tau_d$. However, we will refine this somewhat by using \eqref{cSolution} to find the time $\tau_{curv}$ for which $C=C_{curv}$; i.e., when $\varepsilon(\tau_{curv})=1$:
\begin{equation}
\label{tauCurv}
\tau_{curv}=\frac{\ln2}{3}\,\tau_d\approx0.23\,\tau_d.
\end{equation}

Note that $\tau_{curv}$ is completely independent of the initial particle energy $\varepsilon_0$. This means that the charged particle, which initially has an energy $\varepsilon_0\gg1$, loses a large fraction of this energy over the time $\tau_{curv}$, even if its initial energy is very large. The particle will possess the energy $\varepsilon=1$ at time $t=\tau_{curv}$, after which there will occur the usual exponential decay with time constant $\tau_d$. It is not surprising that the
dependence $\varepsilon(t)$ [see \eqref{curvLargeTimeSolution}] also does not contain the initial particle energy at times $t\gtrsim\tau_d$. This is due to the fact that $\tau_{curv}$ is roughly a factor of four smaller than $\tau_d$, and $\varepsilon$ becomes equal to unity over a time $\tau_{curv}$, independent of $\varepsilon_0$. Over long times, the relation $\varepsilon(\tau_{curv})=1$ begins to play the role of an initial condition for the further exponential decay, as is shown by \eqref{curvLargeTimeSolution}.

All the results we have obtained by averaging \eqref{DLEforUltrarelativisticOscillations} are valid when the adiabatic approximation \eqref{adiabaticCondition} is satisfied. As follows from \eqref{averagedDLEforUltrarelativisticOscillations}, this condition yields an upper limit for the amplitude of the oscillations of the charged particle:
\begin{equation}
\label{amplitudeRestriction}
A\ll A_{\max}=\left(\frac{945}{128\alpha}\right)^{1/7}\omega^{-6/7}\rho^{2/7}
\approx2.6\;\omega^{-6/7}\rho^{2/7}.
\end{equation}
The condition for the adiabatic approximation can also be interpreted as follows. The oscillation period should clearly be smaller than the characteristic decay time for the energy of the charged particle. At first glance, it seems that we should take the exponential decay time $\tau_d$ or the time scale $\tau_{curv}$ for curvature losses as a characteristic time, but this is not so. We must take the more rapid time scale for the energy variations, which, in our case, is the time $\tau_p$ for the transition to a power-law decay. Indeed, if we write the condition
\begin{equation}
\label{equivalentAdiabaticCondition}
T\ll\tau_p,
\end{equation}
the resulting upper limit on the oscillation amplitude coincides fully with the condition \eqref{amplitudeRestriction}. As a consequence, as a criterion for applicability of the adiabatic approximation, we can use any of conditions \eqref{adiabaticCondition}, \eqref{amplitudeRestriction}, or \eqref{equivalentAdiabaticCondition} with equal success.

Let us now estimate to order of magnitude all the quantities introduced above that characterize the time dependence of the energy of a charged particle oscillating near the force-free surface. We will first estimate the energy $C_{curv}$:
\begin{equation}
\label{simCcurv}
C_{curv}\sim\left(\frac{BR^2}{R_L}\right)^{1/3}.
\end{equation}
For a characteristic surface magnetic field $B\sim0.01-0.1$, neutron-star radius $R\sim10^{17}$, and light-cylinder radius $R_L\sim10^{19}{-}10^{20}$ (we will use these values for all our estimates below), $C_{curv}$ is, to order of magnitude, $10^{4}$. We can also introduce the corresponding amplitude
\begin{equation}
\label{simAcurv}
A_{curv}=\frac{\sqrt{2C_{curv}}}{\omega}\sim\left(\frac{RR_L}{B}\right)^{1/3},
\end{equation}
which is $A_{curv}\sim10^{12}{-}10^{13}$ ($\sim1$~m in dimensional units), for the same magnetic field, neutron-star radius, and light-cylinder radius.

The time constant for the exponential decay,
\begin{equation}
\label{simTauD}
\tau_d\sim\frac{R_L}{\alpha B},
\end{equation}
has characteristic values $10^{22}{-}10^{24}$ ($\sim10{-}10^3$~s in dimensional units). The time $\tau_{curv}$ has the same order of magnitude, since relation \eqref{tauCurv} is satisfied. Note that the time $\tau_d$ exceeds the characteristic rotational periods $P$ for pulsars. However, if the field becomes comparable to the critical fields, $B\sim1$, as is true for magnetars, the time $\tau_d$ can be even smaller than the rotational period of the neutron star. Note, by the way, that the ratio $\tau_d/P$ depends only on the surface magnetic field $B$, not on $P$.

The order of magnitude of the time $\tau_p$ cannot be fixed as well, since it depends on the initial particle energy $\varepsilon_0$. However, we can easily find the range in
which $\tau_p$ lies. The upper limit of this range is given by the condition $\varepsilon_0\gg1$, or equivalently $C_0\gg C_{curv}$, which is the only case it makes sense to introduce the concept of the time $\tau_p$. This means that, in any case, $\tau_p\ll\tau_d$. We will determine a lower limit for the range of $\tau_p$ values by taking the maximum possible value of the energy $C_0$. It is obvious that this value is clearly less than the maximum attainable Lorentz factor $\gamma_0$ that the particle will possess far from the force-free surface. Using the fact that $\gamma_0\sim10^{8}$ and assuming $\varepsilon_0\sim\gamma_0/C_{curv}\sim10^{4}$, we find that the inequality $\tau_p\ggg10^{-12}\tau_d$ is always satisfied. This lower limit is made with some reserve, since, in reality, the oscillating particle cannot possess an energy $C_{curv}\sim10^8$: if the particle possessed such an energy far from the force-free surface, its energy after its capture would be somewhat lower, since the particle loses some of its energy in its approach to the force-free surface.

Let us also find to order of magnitude an upper limit $A_{\max}$ for the oscillation amplitude $A$ such that adiabatic approximation \eqref{adiabaticCondition} is still applicable. Using \eqref{amplitudeRestriction}, we obtain
\begin{equation}
\label{simMaxAmplitude}
A_{\max}\sim\left(\frac{R^2R_L^3}{B^3}\right)^{1/7}.
\end{equation}
Thus, $A_{\max}\sim10^{13}{-}10^{14}$ (of the order of several tens of meters in dimensional units). This amplitude corresponds to the oscillation period $T_{\max}\sim A$, which corresponds to $T_{\max}\sim10^{-8}{-}10^{-7}$~s and $\nu\sim10{-}100$~MHz in dimensional units. We can easily find the maximum energy of the oscillations using the amplitude $A_{\max}$:
\begin{equation}
\label{simCmax}
C_{\max}\sim\left(\frac{BR^4}{R_L}\right)^{1/7},
\end{equation}
which is $C_{\max}\sim3\times10^{6}{-}10^7$. This enables us to refine our lower limit for $\tau_p$. Taking as the maximum energy $\varepsilon_0$ the ratio $C_{max}/C_{curv}\sim10^2-10^3$, we obtain the more realistic lower limit $\tau_p\gg10^{-9}\tau_d$.

Finally, let us consider the energy losses of the charged particle in the case of non-relativistic oscillations, which is realized when the oscillation amplitude becomes less than $l_{nro}\simeq1/\omega$. The non-relativistic Dirac–Lorentz equation has the form
\begin{equation*}
\label{nonRelOscEq}
\frac{d^2x'_1}{dt^2}=\frac{2}{3}\alpha\frac{d^3x'_1}{dt^3}-\omega^2x'_1,
\end{equation*}
where the term containing the third time derivative of the coordinate is the well known expression for the force of radiative friction. The corresponding characteristic equation has three roots, one of which is real and positive and the other two of which are complex conjugates. The positive root must be thrown out, since it exhibits the effect of self-acceleration of the particle. The time dependence of the coordinate has the form
\begin{equation*}
\label{nonRelSolution}
x'_1=A_0e^{-t/\tau_{nro}}\cos\omega t,
\end{equation*}
where $A_0$ is the initial amplitude and the decay time constant is
\begin{equation*}
\label{tauDamping}
\tau_{nro}=\frac{3}{\alpha\omega^2}.
\end{equation*}
As expected, to order of magnitude, this coincides with the decay time constant $\tau_{d}$ for the case of ultra-relativistic oscillations.

\section{CAPTURE OF CHARGED PARTICLES}

In \citep{Istomin1Sob'yanin2008}, we found that a charged particle (electron or positron) far from the force-free surface moves in such a way that its Lorentz factor is determined only by its coordinates, and takes on values $\gamma_0\sim10^8$. However, close to the force-free surface, the electric field becomes small and the time for variation of the particle’s energy, $\tau_0=3\rho^2/8\alpha\gamma_0^3$, becomes comparable to the characteristic distance (in dimensionless units) for variations in the field—in the case considered, the distance to the force-free surface. This distance $l_c$ is determined in a self-consistent way from the relation $\tau_0=l_c$; the longitudinal electric field $E_\parallel$, on which $\tau_0$ depends, must be expressed in terms of the distance $l_c$, measured from the force-free surface along the magnetic-field line along which the particle moves. We will use a linear approximation, assuming that $l_c$ is sufficiently small that
\begin{equation}
\label{linearizedE}
E_\parallel(l)=-\omega^2l,
\end{equation}
where we have formally introduced $\omega^2=-dE_\parallel/dl$. Here, the derivative $dE_\parallel/dl$ must be taken at the point where the magnetic-field line along which the particle moves intersects the force-free surface. It is easy to see that $\omega$ is the frequency of non-relativistic oscillations of the particle. Further, we immediately obtain the capture length
\begin{equation*}
\label{captureLength}
l_c=\left(\frac{3}{512\alpha}\right)^{1/7}
\omega^{-6/7}\rho^{2/7}\approx\omega^{-6/7}\rho^{2/7}.
\end{equation*}

Note that the capture length coincides to within a multiplicative coefficient with the maximum oscillation amplitude $A_{\max}$ \eqref{amplitudeRestriction}, determined by the adiabatic approximation condition. This suggests a transition from ultra-relativistic, quasi-stationary motion of the particle far from the force-free surface, determined by the energy-balance condition, to oscillatory motion of the particle near the force-free surface. Qualitative reasoning is sufficient to provide more insight into this. The capture length $l_c$ is nearly a factor of three smaller than the amplitude $A_{\max}$. This suggests that a particle immediately makes a transition to the adiabatic oscillation regime \eqref{adiabaticCondition} upon its capture. To verify this, we will estimate the capture amplitude $A_c$---the maximum distance a particle deviates from the force-free surface when it passes through this surface. We suppose here that the charged particle was initially created far from the force-free surface, began moving toward this surface, and overshot the surface upon its first intersection with it. Let us consider a particle that has not yet intersected the force-free surface, located a distance $l_c$ from this surface. We can estimate the Lorentz factor $\gamma_c$ of the particle at this point using \eqref{linearizedE} and setting $l=l_c$:
\begin{equation*}
\label{captureGamma}
\gamma_c=4\,\omega^2l_c^2.
\end{equation*}

At the same time, assuming the condition \eqref{adiabaticCondition} is satisfied, we can calculate the first integral $C$, setting $\gamma=\gamma_c$ and $x'_1=l_c$:
\begin{equation*}
\label{captureC}
C=\frac{9\,\omega^2l_c^2}{2}.
\end{equation*}
We then have
\begin{equation}
\label{captureA}
A_c=3\,l_c\approx2.9\,\omega^{-6/7}\rho^{2/7}.
\end{equation}
The capture amplitude $A_c$ essentially coincides with the maximum amplitude of adiabatic oscillations $A_{\max}$ \eqref{amplitudeRestriction}. Consequently, we can take a charged particle that has passed through the force-free surface and traveled beyond it a distance $A_c$ to be captured, after which it begins to oscillate in the adiabatic regime that we analyzed in detail above. For oscillations with an initial amplitude $A_c$, the time $\tau_p$ reaches its lower limit, $\tau_p\sim(10^{-9}-10^{-7})\tau_d$.

For completeness, let us find the non-relativistic oscillation frequency $\omega$. Let us turn to \eqref{nonrelativisticOscillationsFrequency}, which determines $\omega^2$. Calculation of the frequency requires computation of the scalar product $\mathbf{B}\cdot\nabla(\mathbf{E}\cdot\mathbf{B})$. In principle, there is no hindrance to calculating this product directly using the expression for the operator $\nabla$ in spherical coordinates. However, we will use a somewhat different approach that enables us to obtain the desired expression in a more compact form. We introduce the unit vectors $\mathbf{e}_r=\mathbf{r}/r$, $\mathbf{e}_m=\mathbf{m}/m$, and $\mathbf{e}_n=\partial\mathbf{e}_m/\partial\theta_m$, where $\mathbf{r}$ is the radius vector, $\mathbf{m}$ the magnetic moment, and $\theta_m$ the angle between the
magnetic axis $\mathbf{e}_m$ and the rotational axis $\mathbf{e}_{\Omega}=\mathbf{\Omega}/\Omega$ of the neutron star. We can then determine the polar angles $\theta$, $\theta'$, and $\theta''$ using the relations $\cos\theta=\mathbf{e}_r\cdot\mathbf{e}_{\Omega}$, $\cos\theta'=\mathbf{e}_r\cdot\mathbf{e}_m$, and $\cos\theta''=\mathbf{e}_r\cdot\mathbf{e}_n$ and the corresponding azimuthal angles $\varphi$, $\varphi'$, and $\varphi''$. Recall that the dipolar magnetic field is potential: $B=-\nabla\Phi$. Here, we have introduced the magnetic potential $\Phi=m\cos\theta'/r^2$, which is a harmonic function: $\triangle\Phi=0$. This enables us to write the desired scalar product as follows:
\begin{equation*}
\label{dotProductInPotentialForm}
\mathbf{B}\cdot\nabla(\mathbf{E}\cdot\mathbf{B})=
\frac{1}{2}\Bigl[\Phi\,\triangle(\mathbf{E}\cdot\mathbf{B})
-\triangle(\Phi\;\mathbf{E}\cdot\mathbf{B})
\Bigr].
\end{equation*}
The calculation itself is carried out by expanding the terms in the functions $P^m_n(\cos\theta)\,e^{im\varphi}/r^l$, where $P^m_n(\cos\theta)$ are associated Legendre polynomials, and further calculating the Laplacian of each of these functions:
\begin{equation*}
\label{LegendreLanlacian}
\triangle\left(\frac{1}{r^l}P^m_n(\cos\theta)\,e^{im\varphi}\right)=
\frac{1}{r^{l+2}}P^m_n(\cos\theta)\,e^{im\varphi}\bigl[l(l-1)-n(n+1)\bigr].
\end{equation*}
In place of the usual angles $(\theta,\varphi)$, we can take the angles $(\theta',\varphi')$ in a spherical coordinate system with the polar axis $\mathbf{e}_m$, or the angles $(\theta'',\varphi'')$ in a coordinate system with the polar axis $\mathbf{e}_n$. Without presenting the computations themselves, which are nevertheless cumbersome, we can write the final result as
\begin{equation}
\label{dotProductFinalExpression}
\mathbf{B}\cdot\nabla(\mathbf{E}\cdot\mathbf{B})=
\frac{km^3}{r^9}\cos\theta'\left[\Bigl(1-\frac{R^2}{r^2}\Bigr)9\cos\theta''\sin\theta_m
+\frac{R^2}{r^2}4\cos\theta\left(11\cos^2\theta'+3\right)\right].
\end{equation}

Substituting \eqref{dotProductFinalExpression} into \eqref{nonrelativisticOscillationsFrequency} yields for the square of the non-relativistic oscillation frequency
\begin{equation}
\label{nonrelativisticFrequencyAtFFS}
\omega^2=-\frac{km}{R^3}\left(\frac{R}{r_{ffs}}\right)^5 4\cos\theta\cos\theta'\,
\frac{2\cos^2\theta'+3}{3\cos^2\theta'+1},
\end{equation}
where $r_{ffs}$ is determined by the equation for the force-free surface,
\begin{equation*}
\label{FFSequation}
r_{ffs}^2=R^2\left(1-4\frac{\cos\theta\cos^2\theta'}{\sin\theta_m\cos\theta''}\right).
\end{equation*}
Here, we have used the relation for the square of the magnetic field $B^2=(3\cos^2\theta'+1)\,m^2/r^6$. Let us also present an expression for the radius of curvature of the field lines:
\begin{equation}
\label{curvatureRadius}
\rho=\frac{r\,(1+3\cos^2\theta')^{3/2}}{3\sin\theta'(1+\cos^2\theta')}.
\end{equation}
If we must calculate $\rho$ at some point on the force-free surface, we must set $r=r_{ffs}$ in this formula.

As a rule, in the entire treatment above, we have not been concerned at all with what type of particle is oscillating near the force-free surface---an electron or a positron. We have assumed only that the sign of the charged particle is such that the force exerted on the particle by the longitudinal electric field $\mathbf{E}_\parallel$ will act to return it toward the force-free surface. Only in this case is it possible to speak of oscillations of the particle. It is now straightforward to refine this question. All the expressions obtained above assumed that the oscillating particle had a positive charge. Consider formula \eqref{nonrelativisticFrequencyAtFFS} for the non-relativistic oscillation frequency. When $\omega^2>0$, we have a positron oscillating with the frequency $\omega$. Accordingly, an electron cannot oscillate under these conditions, because the electric force will repel it from the force-free surface. If we formally suppose that $\omega^2<0$, then, on the contrary, oscillations of electrons will be occur. The non-relativistic oscillation frequency for an electron will then be $\sqrt{-\omega^2}$, and we can take the sign of the oscillating charged particle to coincide with the sign of $\omega^2$.

As follows from \eqref{nonrelativisticFrequencyAtFFS}, the non-relativistic oscillation frequency vanishes if and only if $\theta=\pi/2$ or $\theta'=\pi/2$. The first equality corresponds to the equator of the neutron star, and the second to the star’s magnetic equator. The equator and magnetic equator separate the force-free surface into regions in each of which oscillations of particles of only one sign can occur. The sign of the charged particles changes when passing through the line formed by the non-isolated equilibrium positions. Simultaneously, the equalities $\theta=\pi/2$ and $\theta'=\pi/2$ define a line at each point of which $\omega=0$. However, the signs of the oscillating charges coincide on the open sheets of the force-free surface, which are divided by this line. If $\theta_m<\pi/2$, positrons will accumulate in the arched parts of the force-free surface, and electrons in its open sheets. Note that $\omega^2$ depends on the wavenumber $k$, and therefore also on its sign. If $\theta_m>\pi/2$, we must formally consider the case when the angle between the axes of an inclined rotator is $\pi-\theta_m$, also changing the sign of the angular velocity $\Omega$. Consequently, in this case, the signs of the charges will reverse, so that electrons accumulate in the arched parts and positrons in the open sheets of the force-free surface.

\section{PARTICLE TRAJECTORIES AT THE FORCE-FREE SURFACE}

The motion of a particle captured by the force-free surface is the sum of the drift motion of some driving center that moves exactly along the force-free surface and the oscillatory motion of the particle about this driving center. Let us find the exact trajectories of the driving center in the rotating coordinate frame.

Recall that the velocity of the driving center, $\mathbf{v}'_0$, can be represented as the sum of the drift velocity $\mathbf{v}'_\perp$ perpendicular to the magnetic field and the longitudinal velocity $v'_\parallel\mathbf{b}$. The velocity $\mathbf{v}'_0$ itself always lies in a plane tangent to the force-free surface. However, this velocity possesses another important property. It is easy to see that
\begin{equation}
\label{vEeffOrtogonality}
\mathbf{v}'_0\cdot\mathbf{E}^{eff}=0.
\end{equation}
Here, we have introduced the effective electric field $\mathbf{E}^{eff}=\mathbf{E}+\mathbf{v}_{tr}\times\mathbf{B}$, where $\mathbf{E}$ and $\mathbf{B}$ are the Deutsch electric and magnetic fields and $\mathbf{v}_{tr}=\mathbf{\Omega}\times\mathbf{r}$ is the translational velocity. We can see that the velocity vector $\mathbf{v}'_0$ of the driving center is always orthogonal to the effective electric field $\mathbf{E}^{eff}$. The reason for this is that, first, the drift velocity $\mathbf{v}'_\perp$ is orthogonal to the vector $\mathbf{E}^{eff}$ and, second, the driving center always remains on the force-free surface during its motion, $\mathbf{E}^{eff}\cdot\mathbf{B}=0$, so that the longitudinal velocity component is also orthogonal $\mathbf{E}^{eff}$.

Let us consider the problem of finding an equation for the surface at each point of which the effective electric field $\mathbf{E}^{eff}$ is normal to the surface. We will first consider the inverse problem. Suppose we have some two-dimensional surface specified by the equation $\xi(r,\theta,\varphi)=\mathrm{const}$. Assuming that the function $\xi$ is differentiable, we introduce the vector field $\nabla\xi$. Then, as is well known, this vector field is orthogonal to the surfaces of equal values of the function $\xi$. In
particular, the gradient $\nabla\xi$ at an arbitrary point lying on the specific surface $\xi=\mathrm{const}$ is orthogonal to this surface. Let us now return to the original problem. If there exists a function $\xi$ such that
\begin{equation}
\label{EeffAsNablaXi}
\mathbf{E}^{eff}=\nabla\xi,
\end{equation}
the collection of surfaces that are orthogonal to the vector field $\mathbf{E}^{eff}$, are specified by surfaces of equal values of the function $\xi$. This function exists and is given by the expression
\begin{equation}
\label{potentialXi}
\xi=kR^2\frac{m}{r^3}\left[\left(\cos\theta_m-\cos\theta\cos\theta'\right)
\left(\frac{r^2}{R^2}-1\right)+\frac{2}{3}\cos\theta_m\right].
\end{equation}
As above, here, $(r,\theta,\varphi)$ are spherical coordinates, $R$ the radius of the neutron star, $k=1/R_L$ the wavenumber, $m$ the modulus of the magnetic moment, $\theta_m$ the angle between the rotational and magnetic axes, and $\theta'$ the polar angle relative to the magnetic axis. The potential $\xi$ is determined to within an arbitrary real constant, which we do not write here in the interest of brevity. Expression \eqref{potentialXi} itself was obtained via direct integration of \eqref{EeffAsNablaXi}. Equation \eqref{EeffAsNablaXi} is most easily verified directly. In contrast to $\mathbf{E}^{eff}$, in general, the proper electric field $\mathbf{E}$ is not a potential field, and also contains a vortex component. The equation of the surfaces orthogonal to the vector field $\mathbf{E}^{eff}$ has the form
\begin{equation}
\label{levelSurfaceOfXi}
\xi=\mathfrak{C},
\end{equation}
where $\mathfrak{C}=\mathrm{const}$ is an arbitrary real number if \eqref{levelSurfaceOfXi} has a solution.

First let a particle be located at some point $\mathbf{x}'_0$ of the force-free surface. This point can be substituted into the corresponding number $\mathfrak{C}_0=\xi(\mathbf{x}'_0)$, calculated using \eqref{potentialXi}. Since the velocity of the particle $\mathbf{v}'_0$ lies in a plane tangent to the force-free surface and is orthogonal to the vector $\mathbf{E}^{eff}_0$ at the point $\mathbf{x}'_0$, the velocity lies in a plane tangent to the equipotential surface $\xi=\mathfrak{C}_0$. Thus, the velocity vector $\mathbf{v}'_0$ is tangent to the curve intersecting the force-free surface and the equipotential surface $\xi=\mathfrak{C}_0$ at the point $\mathbf{x}'_0$. This means that the particle moves along this intersection, as can easily be verified by calculating the total time derivative of the potential $\xi$. The potential itself is clearly time independent, $\partial\xi/\partial t=0$, so that its total derivative has the form $d\xi/dt=\mathbf{v}'_0\cdot\nabla\xi$. Recalling that the effective electric field $\mathbf{E}^{eff}$ is a potential field [see \eqref{EeffAsNablaXi}] and that the velocity is orthogonal to $\mathbf{E}^{eff}$ [see \eqref{vEeffOrtogonality}], we immediately obtain $d\xi/dt=0$. Consequently, the potential $\xi$ is conserved along the trajectory of the particle---more precisely, along the trajectory of the driving center---and plays the role of an integral of the motion. We conclude that, as the driving center moves along the force-free surface, it also moves along the equipotential surface specified by the potential $\mathfrak{C}_0$ at the initial point $\mathbf{x}'_0$. As a consequence, the trajectory of the driving center on the force-free surface is determined by specifying the coordinate of only one point through which we construct the trajectory, and not necessarily the initial point.

By virtue of the above, the task of finding the trajectory of the driving center reduces to finding the intersection of the force-free surface and the collection of equipotential surfaces \eqref{levelSurfaceOfXi}. We introduce the function $\zeta(\theta,\varphi)=\xi(r_{ffs}(\theta,\varphi),\theta,\varphi)R/km$,
where $r_{ffs}(\theta,\varphi)$ is the equation for the force-free surface and the potential $\xi$ is determined by \eqref{potentialXi}. The desired trajectory is then given in implicit form by the equation
\begin{equation}
\label{trajectoryFunction}
\zeta(\theta,\varphi)=\mathfrak{C},
\end{equation}
where, as above, $\mathfrak{C}$ is an arbitrary real number for which \eqref{trajectoryFunction} has a solution.

Figure 1 shows a phase portrait of trajectories in the coordinates $(\theta,\varphi)$ on the surface in the rotating frame for the cases of an inclined and orthogonal rotator. Here and below, we assume that the azimuthal angle corresponding to the magnetic axis is zero. We can see that there are no extended trajectories having their beginning and end points on the magnetic equator. This differs from the results of Jackson \citep{Jackson1978}, who asserts that some trajectories on the force-free surface of an uncharged orthogonal rotator end at the magnetic equator, giving rise to a return drift motion at the stellar surface, which sweeps out charged particles. This could hinder the formation of a force-free magnetosphere. However, Jackson assumed that the velocity of a particle at some point of the force-free surface is equal to the projection of the drift velocity onto the tangent plane at that point. We have elucidated above that this is not the case.

Figure 2 presents a three-dimensional image of the trajectories of the driving center on the force-free surface for the case of an inclined rotator. We introduce for convenience in our further analysis the potentials of the isolated equilibrium points,
\begin{equation*}
\label{isolatedPointsPotential}
\zeta_\pm=\mp\frac{1}{3}\frac{(1\mp\cos\theta_m)^{3/2}}
{(3\pm\cos\theta_m)^{1/2}}
\end{equation*}
and the surface potential
\begin{equation*}
\label{nonisolatedPointsPotential}
\zeta_{R}=\frac{2}{3}\cos\theta_m.
\end{equation*}
At the cupola-like parts of the force-free surface, the trajectories have the form of closed cycles surrounding the equilibrium position. The equilibrium position itself corresponds to the potential $\zeta_-$, and the collection of cycles to the interval of the potential $\zeta_R<\zeta<\zeta_-$. The equilibrium positions at the equator and magnetic equator are characterized by the surface potential $\zeta_R$. Two regions can be distinguished in the phase portrait at each of the two open sheets. The first region is analogous to the region of the arches, and also consists of closed cycles surrounding the equilibrium position. The equilibrium position corresponds to the potential $\zeta_+$ and the collection of cycles to the potential interval $\zeta_+<\zeta<0$. However, the cycles themselves become more elongated and tend to go out to infinity as $\zeta$ approaches zero. The separatrix corresponds to the potential $\zeta=0$. The separatrix itself represents an open trajectory that goes out to infinity, but also encompasses the equilibrium position. The second region in each of the open sheets is formed of a collection of trajectories corresponding to the potentials $0<\zeta<\zeta_R$. None of these trajectories lies entirely within a single sheet. As Fig.~2 shows, all the trajectories pass from one sheet to the other, intersecting the line $\theta=\theta'=\pi/2$ and closing around the neutron star. As $\zeta$ approaches $\zeta_R$, these trajectories come closer to the magnetic equator.

The direction of the trajectories is not difficult to determine. For this, we present the components of the velocity $\mathbf{v}'_\perp$:
\begin{equation*}
\label{driftInRotatingFrame}
\begin{aligned}
v'_{\perp r}&=
\Omega r\sin\theta\,b_r b_\varphi\frac{1}{2}\,\Bigl(1-\frac{R^2}{r^2}\Bigr),\\
v'_{\perp \theta}&=
\Omega r\!\left[\cos\theta\,b_r+\Bigl(1-\frac{R^2}{r^2}\Bigr)\sin\theta\,
b_\theta\right]b_\varphi,\\
v'_{\perp \varphi}&=
-\Omega r\!\left[\Bigl(1-\frac{R^2}{r^2}\Bigr)\sin\theta\,\Bigl(\frac{1}{2}b_r^2+b_\theta^2\Bigr)
+\cos\theta\,b_r b_\theta\right],
\end{aligned}
\end{equation*}
where $b_r$, $b_\theta$, and $b_\varphi$ are the components of the unit vector $\mathbf{b}$. By virtue of the structure of the trajectories, we can consider values of the velocity $\mathbf{v}'_\perp$ only at the points of intersection of the force-free surface and the plane $\varphi=\{0,\pi\}$. Here, $b_\varphi=0$, so that $v'_{\perp r}=v'_{\perp\theta}=0$. At the points considered, the velocity of the driving center fully coincides with $\mathbf{v}'_\perp$, so that $v'_\parallel=0$. To determine the required sign of the component $v'_{\perp\varphi}$, we turn to the trajectories that are infinitely close to the magnetic equator, $\cos\theta'=0$. The expression $1-R^2/r^2$ is quadratic in the small quantity $\cos\theta'$, while $b_r$ is linear in this quantity. Consequently, the component $v'_{\perp\varphi}$ changes its sign at the intersection with the magnetic equator, with $\mathrm{sgn}(v'_{\perp\varphi})=-\,\mathrm{sgn}(b_r\cos\theta)$ in a small vicinity of the magnetic equator. Knowledge of this fact is all that is required to reconstruct the direction of the trajectories everywhere on the force-free surface using the principle of continuity. We have at the points of intersection of the force-free surface and the $\varphi=0$ half-plane $v'_{\perp\varphi}>0$ when $\theta\in(0,\theta_m/2)\cup(\pi/2+\theta_m/2,\pi/2+\theta_m)$ and $v'_{\perp\varphi}<0$ when $\theta\in(\theta_m/2,\theta_m)\cup(\pi/2,\pi/2+\theta_m/2)\cup(\pi/2+\theta_m,\pi)$. We have at the points of intersection of the force-free surface and the $\varphi=\pi$ half-plane  $v'_{\perp\varphi}>0$ when $\theta\in(\pi/2-\theta_m,\pi/2-\theta_m/2)\cup(\pi-\theta_m/2,\pi)$ and $v'_{\perp\varphi}<0$ when $\theta\in(0,\pi/2-\theta_m)\cup(\pi/2-\theta_m/2,\pi/2)\cup(\pi-\theta_m,\pi-\theta_m/2)$.

Thus, in the half-space $\cos\theta>0$ in the arched part of the force-free surface and on the open sheet inside the $\zeta=0$ separatrix, the motion along the trajectories
is clockwise as seen by an external observer (of course, the line of sight passing through the force-free surface is directed along the radius toward the center of the star). In the half-space $\cos\theta<0$ in the arched part of the force-free surface and on the open sheet inside the separatrix, the motion along the trajectories is counter-clockwise as seen by an external observer. On the open sheets outside the separatrix (for trajectories intersecting the line $\theta=\theta'=\pi/2$), the motion passes from the half-space $\sin\varphi<0$ to the half-space $\sin\varphi>0$ when $\cos\theta>0$, and from the half-space $\sin\varphi>0$ to the half-space $\sin\varphi<0$ when $\cos\theta<0$. All the above refers to the case $\theta_m<\pi/2$. If the inclination angle of the rotator is $\chi>\pi/2$, we must treat the rotator as having an inclination $\theta_m=\pi-\chi$, with its polar axis, from which all polar angles are measured, being directed opposite to $\mathbf{\Omega}$, and change the sign of $\Omega$. In Fig.~2, this corresponds to changing the direction of $\mathbf{\Omega}$. It is obvious that the directions of all trajectories will be reversed.

Thus, all trajectories of the driving center on the force-free surface are closed and lie in a finite region, apart from the separatrix $\zeta=0$. As a charged particle moves along its trajectory, the electromagnetic field, and therefore the parameters of the particle’s oscillation, change. However, the driving center moves with a drift velocity $\sim10^{-4}$, and, as follows from \eqref{simMaxAmplitude}, the oscillation frequency exceeds $10$~MHz, even for ultra-relativistic oscillations, so that the particle traverses a distance of the order of several millimeters, appreciably less than $R$, over the oscillation period. As a consequence, the variation of the parameters of the particle oscillation during its motion along the force-free surface is adiabatic. Note that the sign of $\omega^2$ does not change in the motion along the trajectory [see the discussion following \eqref{curvatureRadius}]. This means that, if a particle falls into the force-free surface, it is not able to leave, since, first, the trajectory of its driving center is closed and, second, instability of the particle oscillations along the magnetic field does not develop due to the constancy of the sign of $\omega^2$.

\section{ACCUMULATION OF ELECTRON–POSITRON PLASMA}

Let us estimate the rate of accumulation of electron–positron plasma at the force-free surface. We will consider the situation when the charged particles---electrons and positrons---move in a specified electromagnetic field, playing the role of test charges. This is possible only while the density of charges in the magnetosphere is appreciably less than the Gold\-reich–Julian density $\rho_{GJ}$. However, the rate at which pairs are created in the magnetosphere is constant, so that, with time, particles are accumulated at the force-free surface. Consequently, a region appears in the vacuum magnetosphere near the force-free surface where the charge density becomes comparable to $\rho_{GJ}$ over some finite time. As a result, the entire magnetosphere becomes filled with plasma.

We first consider the initial stage in the restructuring of the magnetosphere from a vacuum state to a state in which it is filled with plasma. We choose an arbitrary point $\mathbf{r}_0$ on the force-free surface and consider a small area on this surface containing this point. The size of this area is assumed to be small compared to $R$, so that the area is virtually flat. With time, charged particles accumulate in this area, forming a symmetrical charge layer.

We will consider the accumulation of plasma only for fairly short times, when the thickness of the formed charge layer is appreciably smaller than $R$. In this case, the plasma density differs appreciably from zero only near the force-free surface, so that there is not yet a global restructuring of the magnetosphere. Consequently, the outer electromagnetic field where the charged layer is located does not change. Let us find the proper electric field of the layer at some distance $z$ from the force-free surface:
\begin{equation}
\label{intrinsicE}
E_l=4\pi\alpha\int\limits_0^z\rho_e(z')\,dz',
\end{equation}
where $\rho_e(z')$ is the volume charge density a distance $z'$ from the force-free surface and the distance $z$ does not exceed $H$---the distance from the force-free
surface to the boundary layer. Here, it is important that the charge layer is symmetrical, $\rho_e(-z')=\rho_e(z')$. Providing an equilibrium state for such a layer requires that each charged particle belonging to the layer not be accelerated along the magnetic-field lines. In other words, the outer longitudinal electric field \eqref{linearizedE} must be fully balanced by the projection of the electric field of the layer \eqref{intrinsicE} onto the direction of the magnetic field. Introducing the angle $\psi$ between the normal to the force-free surface and the direction of the magnetic field,
\begin{equation*}
\label{cosPhi}
\cos\psi=\frac{\mathbf{B}\cdot\nabla(\mathbf{E}\cdot\mathbf{B})}
{B\,|\nabla(\mathbf{E}\cdot\mathbf{B})|}.
\end{equation*}
The relation
\begin{equation}
\label{zeroParallelFieldCondition}
4\pi\alpha\cos\psi\int\limits_0^{l\cos\psi}\rho_e(z')\,dz'=\omega^2l
\end{equation}
is then satisfied for arbitrary $l$. Differentiating both sides of \eqref{zeroParallelFieldCondition} with respect to $l$, we obtain
\begin{equation}
\label{chargeDensity}
\rho_e=\frac{\omega^2}{4\pi\alpha\cos^2\psi}.
\end{equation}

Thus, the charge density $\rho_e$ in the layer does not depend on the distance from the force-free surface. Recall that we are measuring the charge density in units of $e/{^-\!\!\!\!\lambda}^3$. This is determined only by the square of the non-relativistic oscillation frequency $\omega^2$ and the angle $\psi$, whose values are taken at the point $\mathbf{r}_0$ lying in the small area of the force-free surface we are considering. The sign of the charge density $\rho_e$ coincides with the sign of $\omega^2$, in agreement with our conclusions following \eqref{curvatureRadius}. In general, $\rho_e$ does not coincide with $\rho_{GJ}$. This flows from the fact that, if the charge density in the layer were everywhere equal to the Goldreich–Julian charge density, all the particles in the layer would be fully co-rotating, so that they would be stationary in the rotating frame. In the layer, only the longitudinal electric field is zero, while the fields perpendicular component differs from zero, because the electric field of the layer does not necessarily cancel out the perpendicular component of the external field. For example, because the electric field of the charged layer is equal to zero at the point $\mathbf{r}_0$, the total electric field has only a perpendicular component, which gives rise to a drift of the charged particles along the force-free surface, along the trajectories depicted in Fig.~2. This means that the forming charge layer is made up of differentially flowing currents on the force-free surface.

Thus, a charge layer forms near the force-free surface with time, whose thickness $H$ is a function of the point $\mathbf{r}_0$ and the time $t$ that has passed since the start of the filling of the magnetosphere with plasma. It is important that only charges of the same sign as the charge density $\rho_e$ contribute to the value of $\rho_e$ at each point of the force-free surface. Because of this, the layer is fully charge-separated in the initial stage of filling of the magnetosphere, in accordance with our conclusions following \eqref{curvatureRadius}. Consequently, the number density of the particles $n_e$ coincides with the plasma density: $n_e=|\rho_e|$. At each point of the force-free surface, we introduce the current density of the particles $j=dN/dSdt$, which coincides with the electric-current density in dimensionless units. The equation used to find the thickness of the charge layer then has the form
\begin{equation}
\label{hEq}
\frac{\partial H}{\partial t}=\frac{j}{n_e}.
\end{equation}

Equation \eqref{hEq} can be used to estimate the time to fill the entire vacuum magnetosphere with electron-positron plasma. When the thickness of the charge layer $H$ becomes comparable to the characteristic size of the inner magnetosphere, i.e., to the radius of the star $R$, there is a substantial restructuring of the entire magnetosphere. The filling time is then
\begin{equation*}
\tau_f=\frac{Rn_{GJ}}{j}.
\end{equation*}
Here, we have taken into account that the plasma density in the layer is of the order of the Goldreich-Julian density ($n_e\simeq n_{GJ}$), as follows from \eqref{chargeDensity}. The current density of the particles incident onto the force-free surface $j$ is determined by the mechanism creating pairs in the magnetosphere. Gamma-rays capable of producing electron–positron pairs in the magnetosphere magnetic field can both be incident from outside, as cosmic rays, and be produced from soft thermal photons radiated by the stellar surface that are Compton scattered by energetic particles. The main question here is how many pairs are produced by a single photon with energy exceeding 1~MeV. Since the created electrons and positrons are very rapidly accelerated to energies $\gamma_0$, there is a chain multiplication of pairs. The number of created pairs per primary photon depends exponentially on the characteristic size of the inner magnetosphere: $\exp{\mu}$, where $\mu=R/l_\Sigma$. Here, $l_\Sigma$ is the total mean free path of an energetic particle, including both curvature radiation and pair creation.

Studies of processes associated with the stationary creation of plasma in the polar magnetosphere \citep{GurevichIstomin1985} indicate that the mean-free path $l_\Sigma$ is of the order of 100~m. This means that the multiplication coefficient can reach gigantic values of the order of $\exp{100}\simeq 10^{43}$. However, in view of the presence of the exponential factor, resolving this question for the case of a vacuum magnetosphere requires a more careful analysis, which falls outside the framework of this paper.

\section{CONCLUSION}

We have shown that as a charged particle that is initially accelerated to a Lorentz factor $\gamma_0\sim10^8$ approaches the force-free surface, the quasi-stationary condition is violated, and the particle passes through the force-free surface with a Lorentz factor $C_{\max}\sim10^6-10^7$ \eqref{simCmax}. After passing through the force-free surface, the particle deviates from this surface by a distance $A_c\sim A_{\max}\sim10-100$~m \eqref{amplitudeRestriction}, \eqref{simMaxAmplitude}, \eqref{captureA}. Further, the particle undergoes adiabatic ultra-relativistic oscillations at a frequency $\nu\sim10-100$~MHz \eqref{simMaxAmplitude}. These oscillations decay due to energy losses to radiation, and their frequency grows. The energy of the oscillations falls first linearly, then, over a time $\tau_p\sim10^{-8}-10^{-4}$~s [see \eqref{tauP} and the comments following \eqref{simCmax} and \eqref{captureA}], according to a power-law
decay, due to losses to curvature radiation. Over times of the order of $\tau_{curv}\sim\tau_d\sim10-1000$~s \eqref{tauD}, \eqref{tauCurv}, \eqref{simTauD}, the decay becomes exponential, with the time constant $\tau_d$ \eqref{tauD}, with the main contribution to the energy losses now made by bremsstrahlung. During the change in the decay regime, the particle possesses a Lorentz factor $C_{curv}\sim10^4$ \eqref{Ccurv}, \eqref{simCcurv}, and the oscillation amplitude is $A_{curv}\sim1$~m \eqref{simAcurv}. Further, the ultra-relativistic oscillations of the particle continue to exponentially decay. When the amplitude $l_{nro}\sim1$~cm is reached, the oscillations become non-relativistic and harmonic, and their frequency is $\nu\sim1-10$ GHz. Simultaneous with this oscillatory motion, the particle undergoes regular motion along the force-free surface (Fig.~2). As a rule, the trajectory of the driving center is closed in the rotating frame, and its velocity is of the order of the drift velocity.

The gradual accumulation of particles at the force-free surface leads to the formation of a fully charge-separated layer of electron–positron plasma. The volume density of this plasma depends on the coordinates $(\theta,\varphi)$ on the force-free surface, but not on the layer thickness. This layer has a sharp boundary, and the plasma density in the layer is comparable to the Goldreich–Julian density. In the case of a constant rate of creation of pairs in the magnetosphere, the layer thickness grows monotonically. Over a finite time, the layer thickness becomes comparable to the radius of the neutron star, accompanied by a restructuring of the magnetosphere. This is the characteristic time for filling of the magnetosphere with plasma. The efficient acceleration of the particles gives rise to a chain of creations of secondary pairs. The resulting high multiplication coefficient for pair creation may be compensated for by the smallness of the flux of cosmic-ray photons. According to the observations \citep{StrongEtal2005,SizunEtal2006}, the diffuse background of Galactic photons $j_{ph}$ having energies exceeding 1~MeV is of the order of $10^{-3}\text{ cm}^{-2}\text{s}^{-1}$. Thus, the ratio $j_{ph}/cn_{GJ}\simeq 10^{-25}$ is very small. However, this does not hinder the rapid filling of a pulsar magnetosphere with electron–positron plasma if the pair-creation multiplication coefficient is sufficiently high.

\section*{ACKNOWLEDGMENTS}

This work was partially supported by the Russian Foundation for Basic Research (project code 08-02-00749-a), the Program of State Support for Leading Scientific Schools of the Russian Federation (grant no. NSh-1738.2008.2), and the Federal Agency for Science and Innovation (state contract no. 02.740.11.0250).
\newpage
\renewcommand{\refname}
\begin{flushright}
\textit{Translated by D. Gabuzda}
\end{flushright}
\newpage
\renewcommand{\figurename}{Fig.}
\begin{figure}
\includegraphics[width=\textwidth]{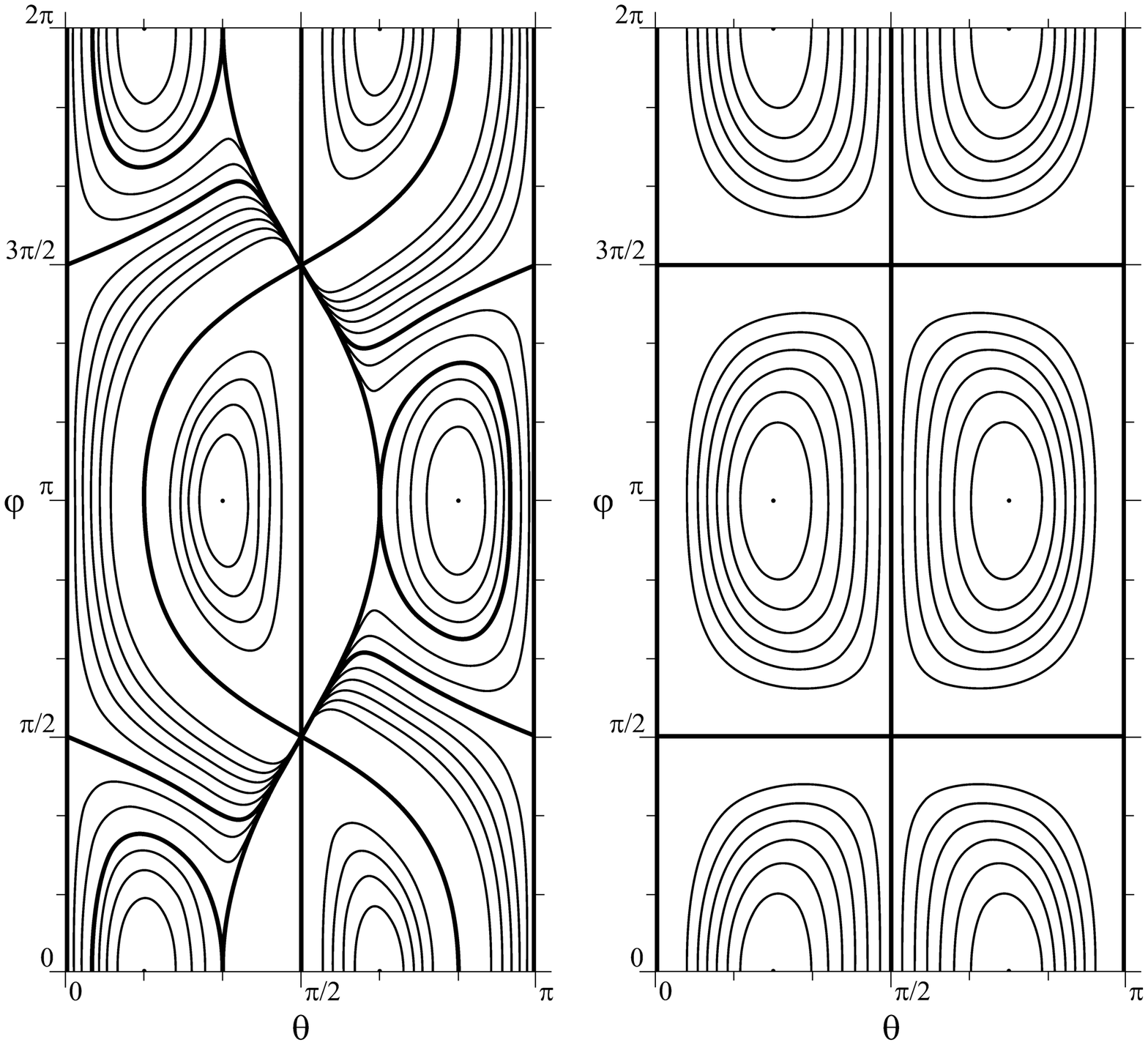}
\caption{Phase portrait of the trajectories of charged particles at the force-free surface in the coordinates $(\theta,\varphi)$ in the rotating frame for rotators whose rotational and magnetic axes are mutually inclined by $\pi/3$ (left) and $\pi/2$ (an orthogonal rotation, right).}
\end{figure}
\newpage
\begin{figure}
\includegraphics[width=\textwidth]{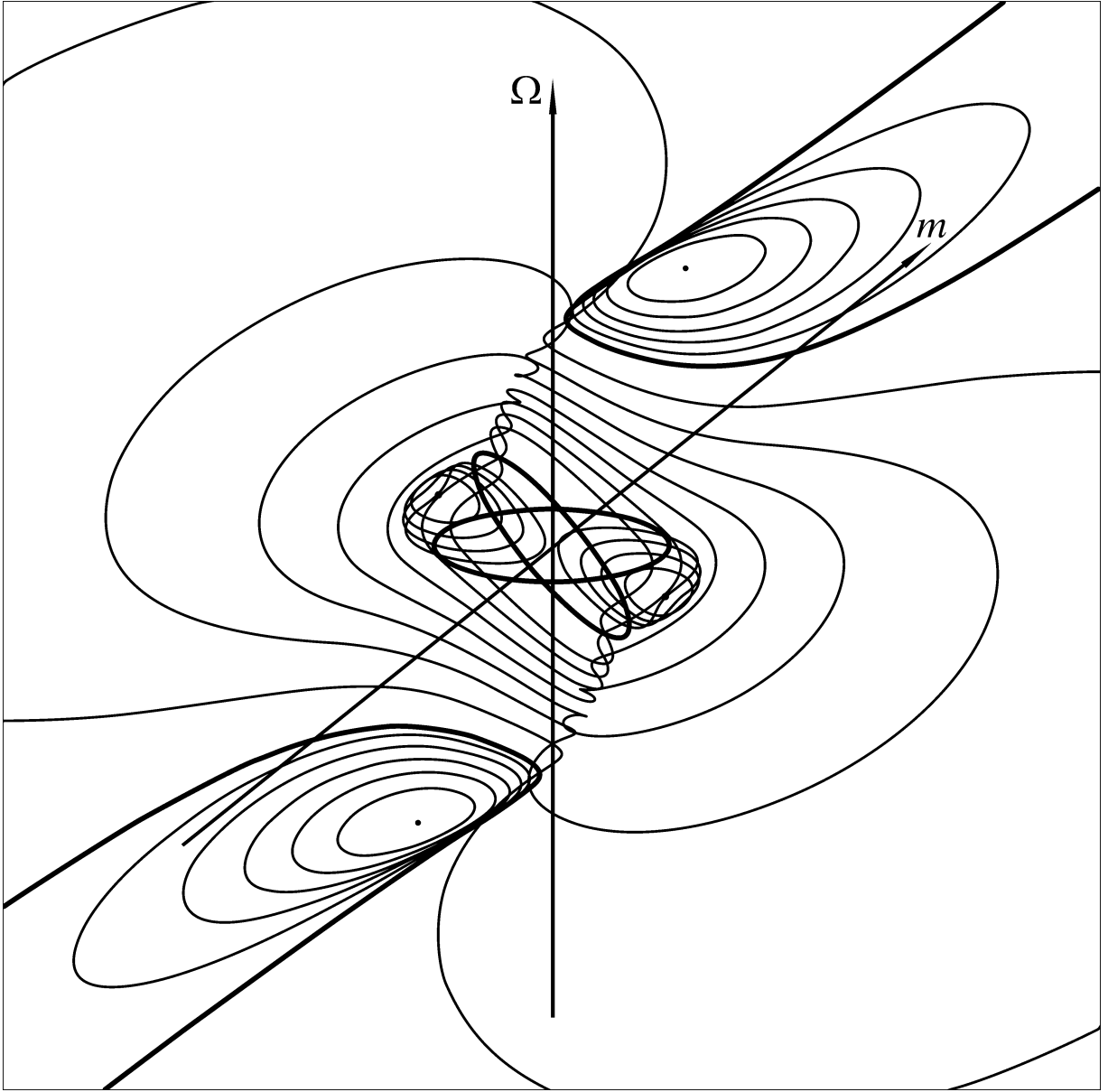}
\caption{Trajectories of charged particles on the force-free surface in the rotating frame for the case of a rotator with its rotational and magnetic axes inclined by $\pi/3$.}
\end{figure}

{\large\centering REFERENCES}
\begin{thebibliography}{99}
\bibitem[Sturrock(1971)]{Sturrock1971}
P. A. Sturrock,
Astrophys. J. \textbf{164}, 529 (1971).
\bibitem[Ruderman and Sutherland(1975)]{RudermanSutherland1975}
M. A. Ruderman and P. G. Sutherland,
Astrophys. J. \textbf{196}, 51 (1975).
\bibitem[Goldreich and Julian(1969)]{GoldreichJulian1969}
P. Goldreich and W. H. Julian,
Astrophys. J. \textbf{157}, 869 (1969).
\bibitem[Gurevich and Istomin(2007)]{GurevichIstomin2007}
A. V. Gurevich and Ya. N. Istomin,
Mon. Not. R. Astron. Soc. \textbf{377}, 1663 (2007).
\bibitem[Kramer et~al.(2006)]{KramerEtal2006}
M. Kramer, A. G. Lyne, J. T. O'Brien, et~al.,
Science \textbf{312}, 549 (2006).
\bibitem[Kramer(2008)]{Kramer2008}
M. Kramer,
AIP Conf. Proc. \textbf{983}, 11 (2008).
\bibitem[Wang et~al.(2007)]{WangEtal2007}
N. Wang, R. N. Manchester, and S. Johnston,
Mon. Not. R. Astron. Soc. \textbf{377}, 1383 (2007).
\bibitem[McLaughlin et~al.(2006)]{McLaughlinEtal2006}
M. A. McLaughlin, A. G. Lyne, D. R. Lorimer, et~al.,
Nature \textbf{439}, 817 (2006).
\bibitem[Istomin and Sob'yanin(2009)]{Istomin1Sob'yanin2008}
Ya. N. Istomin and D. N. Sob’yanin,
Astronomy Reports, \textbf{54}, 338 (2010).
\bibitem[Jackson(1978)]{Jackson1978}
E. A. Jackson,
Astrophys. J. \textbf{222}, 675 (1978)
\bibitem[Gurevich and Istomin(1985)]{GurevichIstomin1985}
A. V. Gurevich and Ya. N. Istomin,
Zh. \'{E}ksp. Teor. Fiz. \textbf{89}, 3 (1985)
[Sov. Phys. JETP \textbf{62}, 1 (1985)].
\bibitem[Strong et~al.(2005)]{StrongEtal2005}
A. W. Strong, R. Diehl, H. Halloin, et~al.,
Astron. Astrophys. \textbf{444}, 495 (2005).
\bibitem[Sizun et~al.(2006)]{SizunEtal2006}
P. Sizun, P. Cass\'{e}, and S. Schanne,
Phys. Rev. D \textbf{74}, 063514 (2006).
\end{thebibliography}
\end{document}